# Hyperscaling relation between the interfacial tension of liquids and their correlation length near the critical point

E. Mayoral[1i*] and A. Gama Goicochea[2ii*]

[1] Instituto Nacional de Investigaciones Nucleares, Carretera México-Toluca s/n, La Marquesa Ocoyoacac, 52750 Estado de México, México

[2] Instituto de Física, Universidad Autónoma de San Luis Potosí, Av. Álvaro Obregón 64, 78000 San Luis Potosí, Mexico

## ABSTRACT

Interfaces involving coexisting phases in condensed matter are essential in various examples of soft matter phenomena such as wetting, nucleation, morphology, phase separation kinetics, membranes, phase coexistence in nanomaterials, etc. Most analytical theories available use concepts derived from mean field theory which does not describe adequately these systems. Satisfactory numerical simulations for interfaces at atomistic to mesoscopic scales remains a challenge. In the present work, the interfacial tension between mixtures of organic solvents and water is obtained from mesoscopic computer simulations. The temperature dependence of the interfacial tension is found to obey a scaling law with an average critical exponent $\mu = 1.23$. Additionally, we calculate the evolution of the correlation length, defined as the thickness of the interface between the immiscible fluids, as a function of temperature and find that it obeys also a scaling law with the average critical exponent being $\nu = 0.67$. Lastly, we show that the comparison of $\mu$ and $\nu$ for these binary mixtures constitutes the first test of Widom's hyperscaling relation between these exponents in 3$d$, expressed as $\mu = \nu(d - 1)$. Based on these values and those for the 3$d$ Ising model it is argued that both systems belong to the same universality class, which opens up the way for the calculation of new scaling exponents.

---

[i] Electronic mail: emayoral@inin.gob.mx
[ii] Electronic mail: agama@alumni.stanford.edu

# 1. INTRODUCTION

The relationship between properties and structure of several physical, chemical and biological systems is important to understand and design new materials in soft condensed matter. Morphologies can be established in terms of uniform or non uniform regions separated by interfaces where the physicochemical properties change abruptly in different thermodynamic conditions, such as changes in temperature, which is a foremost challenge with significant implications. A recent example is in the development of enhanced oil recovery techniques, where multiphase compositions are present and it is important to understand the properties of the interfacial tension to control instabilities when the temperature is modified. Additional processes involving interfaces of novel significance occur at or near critical points which cannot be satisfactorily described by traditional methods. Computation of interfacial properties between coexisting phases is also useful to characterize interfacial phenomena such as wetting, capillary condensation, heterogeneous nucleation, etc. For these reasons among others the study of interfacial tension between immiscible liquids and its dependence on temperature near the critical point is an important ingredient in the study of different soft matter systems.

The concepts of scaling and universality have given rise to the modern theory of critical phenomena which is still a leading research topic in different areas of modern physics and chemistry [1 – 3]. These systems possess two general characteristics at their critical point: *universality*, meaning that they display the same critical behavior, and *scaling,* because in the neighborhood of the critical point the system is invariant under scale transformations. Hence, a useful understanding of systems near their critical point can be achieved through power laws whose properties may be analyzed using dimensional considerations known as scaling laws [3]. The Renormalization Group (RG) is perhaps the most accurate predictive tool to obtain power laws for many systems in close proximity to their critical point [2, 4], since mean field theory is only correct when fluctuations of the order parameter are much smaller than its mean value, which does not occur near the critical point. For this reason the use of equations of state such as that of van der Waals are not an adequate route to obtain critical exponents [5 – 7]. Although RG is a powerful alternative to obtain accurately critical exponents [8, 9], its application to soft matter systems is far from trivial, while on the other hand, computer simulations can solve models highly accurately for a wide variety of complex fluids [10]. Coarse grained simulations are particularly useful for the study of scale – invariant fluids. One of the most popular coarse graining approaches is the so called Dissipative Particle Dynamic (DPD) model [11, 12], which has successfully predicted the behavior of various types of systems [13 - 17]. Scaling properties at constant temperature have been obtained using DPD [18], but to study critical phenomena the temperature dependence must be taken into account fully. Recently we developed a procedure to introduce the temperature dependence in DPD simulations [19], based on the estimation of the temperature dependence of the Flory – Huggins $\chi(T)$ parameter from the calculation of cohesive energy densities of the components at different temperatures using microscopic molecular dynamics (MMD). Here we show how this parametrization of the temperature dependence leads to the correct description of universality and scaling for liquid – liquid systems near the critical point.

## 2. MODELS AND METHODS

We obtain first the scaling exponent of the dependence of the interfacial tension $\sigma$ with the temperature, $T$, for mixtures of immiscible fluids. Such dependence is expected to be given by [7]:

$$\sigma(T) = \sigma_0 (1 - T/T_C)^\mu, \qquad (1)$$

where $\sigma_0$ is a system – dependent constant, $T_C$ is the critical temperature at which the interface becomes unstable, and $\mu$ is a critical exponent, which was found experimentally to be close to 11/9 [20]. More recently, RG and Monte Carlo calculations [8, 9, 21 - 23] yielded the currently accepted value $\mu = 1.26$; using then Widom's hyperscaling relation [7]:

$$\mu = (d-1)\nu, \qquad (2)$$

the value for the exponent $\nu$ can be obtained. Equation (2) relates the exponent $\mu$ with the dimensionality $d$ of the system and with the critical exponent $\nu$ associated with the correlation function, $\xi$. Testing the validity of eq. (2) is the central purpose of this work. To do so we first obtain directly the scaling exponent ($\mu$) from the temperature dependence of the interfacial tension between immiscible liquids using DPD simulations, and compare our predictions with those reported in the literature. At the same time, we extract the critical exponent of the correlation length, $\nu$, from the thickness of the interface between the liquids as the temperature approaches $T_C$. By contrast, most works on this subject have focused on obtaining $\mu$ and then, *assuming* eq. (2), a value for $\nu$ is predicted [8, 9, 21 – 23]. Using the commonly accepted value of the exponent in eq. (1) $\mu = 1.26$ [9], and applying eq. (2) one obtains $\nu = 0.63$ for $d = 3$, which is equal to its expected vale for the 3$d$ Ising model (see, for example, [24] and references therein). We obtain separate predictions for $\mu$ and $\nu$, and compare them with eq. (2); to our knowledge this is the first time eq. (2) has been proved.

The exponent $\nu$ is defined from the temperature dependent correlation length ($\xi$) [1]:

$$\xi = \xi_0 |t|^{-\nu}(1 + \cdots) . \qquad (3)$$

In this expression $t$ is the reduced temperature defined as $t \equiv \frac{T-T_C}{T_C}$. The exponent $\mu$, as defined in eq. (1), is related with $\nu$ as shown in eq. (2) [7]; here only the 3$d$ case is tested, $\mu = 2\nu$. We perform simulations of binary mixtures of water with benzene, dodecane, dodecanol and hexanol at various temperatures. The DPD method is by now well know, therefore we shall be brief here; for a recent review of the method and some of its applications, see [12]. All the DPD thermodynamic properties are obtained from the conservative force acting between any pair of particles $i$ and $j$, which is proportional to a constant $a_{ij}$, while dissipative and random forces keep the temperature fixed [13]. The temperature dependence of $a_{ij}$ is obtained from the calculation of the temperature dependence of the solubility parameters using MMD [19]. The simulations were performed with 4500 DPD particles in a cubic box with $L^* = 11.4$. Previous studies [25, 26] have shown that DPD simulations are relatively insensitive to finite size effects, as a consequence of the short range of the interactions. In fact, the difference between the interfacial tension between two simple DPD fluids in a box with $L^* = 5$ and that with $L^* =$ 11 was found to be smaller than one percent [25]. By contrast, the capillary wave method for the calculation of the interfacial tension is more sensitive to the size of the simulation box [27]. Full details can be found in the Supplementary Information (SI).

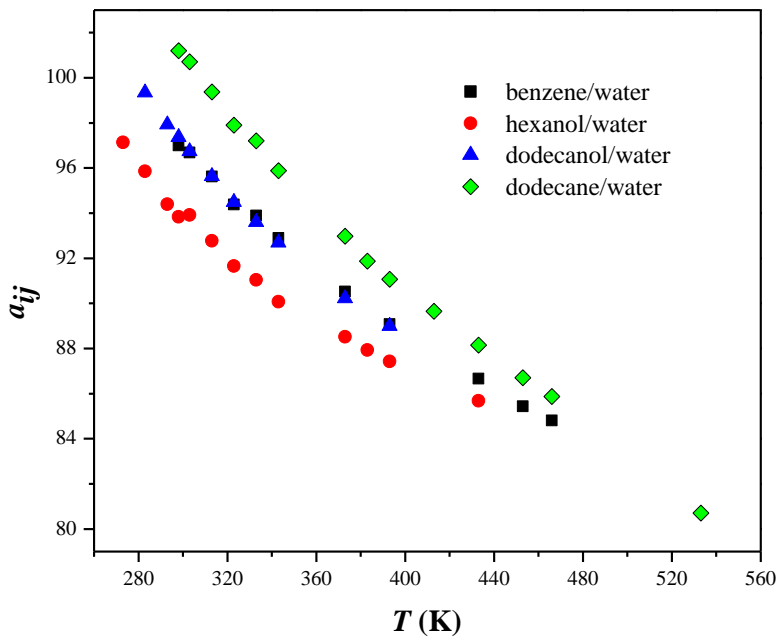

**Figure 1.** DPD conservative interaction parameter $a_{ij}$ in reduced units, at different temperatures for benzene/water, hexanol/water, dodecanol/water and dodecane/water.

## 3. RESULTS AND DISCUSSION

Figure 1 shows the dependence on temperature of the DPD interaction parameter for all the systems studied, which have qualitatively the same behavior as that of mixtures of organic solvents and water reported recently [19]. Using the values of $a_{ij}$ shown in Fig. 1 we ran simulations of each binary mixture at each temperature, to calculate the interfacial tension at that given temperature, $\sigma(T)$. The latter is calculated from the difference between the component of the pressure tensor perpendicular to the interface between the fluids, and the average of the components parallel to the interface. These components are obtained from the virial theorem, which leads to a kinetic contribution (ideal gas term), and one arising from the interactions. For these we used the Irving – Kirkwood method, see SI. The values of $\sigma(T)$ obtained from our simulations are shown in Fig. 2. The constants $\sigma_0$, $\mu$ and $T_C$ were obtained from the best fits of the $\sigma(T)$ vs $T$ data. The value of $\mu$ found from the fitting of our simulation data for the binary mixtures we modeled, $\mu = 1.2 \pm 0.1$, is within the statistical error, in agreement with the expected value [8, 9, 23, 24] $\mu = 1.26$, although the agreement is somewhat less satisfactory for the dodecanol/water mixture (green rhombi in Fig. 2), where we find $\mu = 1.3 \pm 0.3$.

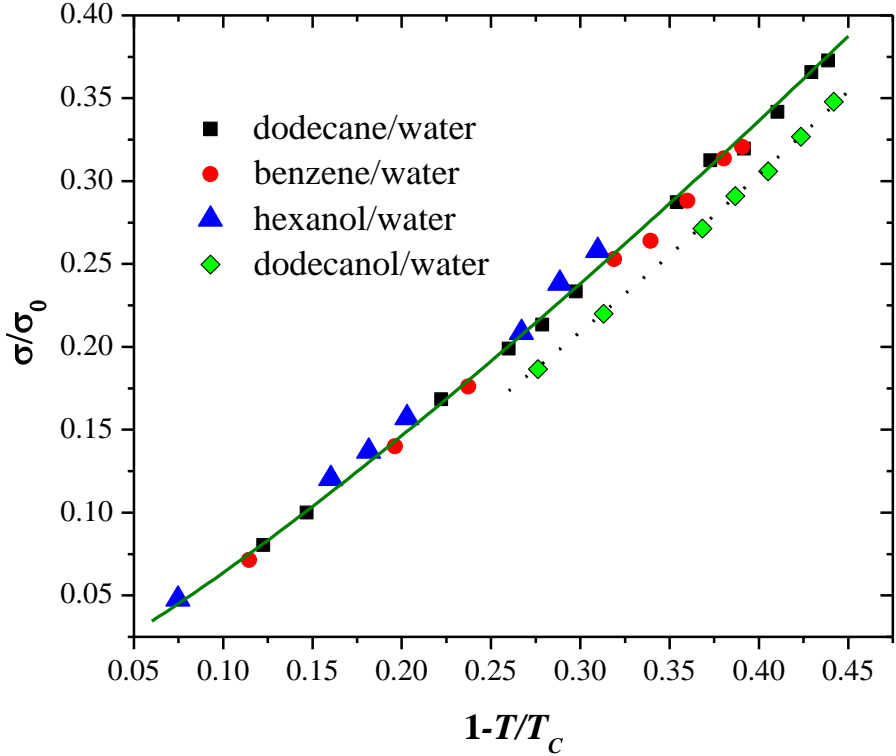

**Figure 2**. Interfacial tension for all the binary mixtures simulated, as a function of temperature, obtained from DPD simulations. The solid line represents the function $\sigma/\sigma_0 = (1 - T/T_C)^\mu$, with $\mu = 1.2$, while the dotted line corresponds to the same function with $\mu = 1.3$.

This is attributed to the fact that dodecanol was mapped as a four – bead linear polymer chain whose "head" and three – bead "tail" have different values of the conservative interaction parameter ($a_{ij}$), to account for the combined effects of having the largest chain among the four organic compounds we studied, and having polar character because of its OH group. The long hydrophobic chain coupled with the polar OH – group in the dodecanol molecule leads to a relatively more abrupt reduction of the interfacial tension at a given temperature than that obtained with any of the other organic compounds we modeled, which accounts for a larger value of $\mu$. This is supported by the density profiles obtained from our simulations, which at a given temperature show that the dodecanol/water interface is the one with the largest width, and this in turn translates as a lower interfacial tension. The results shown in Fig. 2 confirm the validity of eq. 1; however an independent route is required to obtain $\nu$ so that it can be compared with $\mu$ to test the validity of eq. 2. We define the correlation length $\xi$ as the thickness of the interfacial region between the immiscible fluids, which becomes increasingly large as the temperature approaches $T_C$. To determine $\xi$ we have used the time averaged density profile, which leads to an interfacial thickness that contains not only the so – called intrinsic width [28] but also the fluctuations of the interface due to capillary waves. Then, using eq. 3 we obtain the critical exponent $\nu$. For the calculation of $\xi$ we take the point where the density of the fluid has reached its bulk value, as shown in Fig. 3.

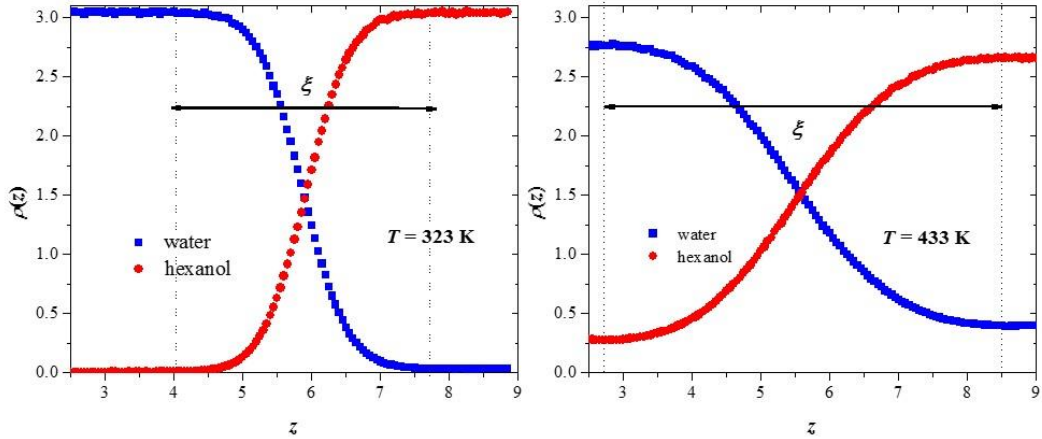

**Figure 3**. Temperature dependence of the thickness ($\xi$) of the interface between hexanol and water as the temperature is increased. The dotted lines indicate the size of the correlation length. The axes are shown in reduced DPD units.

In Fig. 3 we illustrate this procedure for the particular case of the hexanol/water system at two different temperatures, where the thickness of the interface, i.e. the correlation length is indicated by the arrows and the dotted lines. As expected, $\xi$ increases as the $T$ approaches $T_C$. Similar trends are found for the other three systems modeled and the full series of density profiles at various temperatures can be found in the SI. With the data obtained from these trends, and the values of $T_C$ extracted from the fitting shown in Fig.2 we apply eq. 3 to extract a value for $\nu$ from our simulations, without using eq. 2.

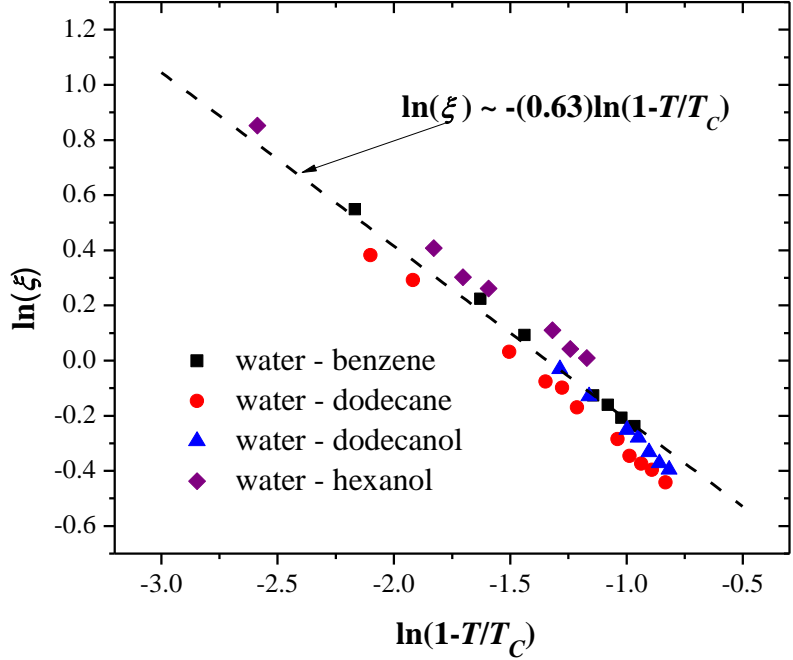

**Figure 4**. Correlation length ($\xi$) dependence on temperature, extracted from the thickness of the interface between the two immiscible liquids (see also Fig. 3). The dotted line represents the scaling of the correlation length expected for the $3d$ Ising model, $\xi \sim (1 - T/T_C)^{-0.63}$, see text for details.

The results from this procedure are shown in Fig. 4. From the best fits for each binary mixture we find $\nu = 0.59 \pm 0.01$ for hexanol/water; $\nu = 0.66 \pm 0.01$ for dodecane/water; $\nu = 0.67 \pm 0.12$ for benzene/water, and $\nu = 0.78 \pm 0.02$ for dodecanol/water. These predictions are all reasonably close to the expected value for the $3d$ Ising model $\nu = 0.63$ [29], with the exception of the dodecanol/water case, for the reasons provided above in the discussion of Fig. 2. Multiplying these values of $\nu$ by two to obtain the predicted $\mu$ obtained from the scaling relation $\mu = 2\nu$ (eq. 2), and comparing with those obtained directly from the prediction of the interfacial tension and shown in Fig. 2 we find that Widom´s hyperscaling relation (eq. 2) is fulfilled (within statistical accuracy) in *all* four cases studied here. Moreover, the average values of the critical exponents found from our simulations, $\langle \mu \rangle = 1.23 \pm 0.15$ and $\langle \nu \rangle = 0.67 \pm 0.04$, are in agreement (within statistical accuracy) with the accepted values for the 3$d$ Ising model, i.e., $\mu$ = 1.26 and $\nu$ = 0.63 [29]. We therefore must conclude that the interfacial tension between the organic fluids and water modeled here belongs to the same universality class as that of the 3$d$ Ising model. Although there are other works where the temperature dependence of the interfacial tension was calculated using methods different from ours [30 – 34], none of them predicted the temperature dependence of the correlation length simultaneously. This is the first test of Widom´s hyperscaling relation that we know of. Based on our results, in 2$d$ we expect $\mu = 1$ since $\nu = 1$ for the 2$d$ Ising model [29].

## 4. CONCLUSIONS

We have shown that the temperature dependent interfacial tension of binary mixtures can be fitted to a universal power law (eq. 1) with an average exponent $\langle \mu \rangle = 1.23 \pm 0.15$. Additionally, the thickness of the interface formed between immiscible fluids is found to be a useful definition of the correlation length, whose dependence on temperature yields an average value for critical exponent $\langle \nu \rangle = 0.67 \pm 0.04$. Both predictions are in agreement with the values of the exponents for the 3$d$ Ising model, therefore both problems belong to the same universality class. Finally, the successful testing of the hyperscaling relation between $\mu$ and $\nu$ shows that the temperature dependent DPD model used here preserves the scale – invariant symmetry and the fundamental scaling properties of mesoscopic systems near their critical point. Critical exponents such as $\mu$ and $\nu$ are fundamental for the understanding of universality in different phenomena to improve applications related with interfacial tension and with the characteristic correlation length of different chemical, biological and physical systems. Lastly, knowing the $\nu$ exponent is also important because it is related with the structure and scaling of polymer chains in a solvent, which leads to special insights into the design of multi-component polymeric materials.

## ACKNOWLEDGEMENTS

AGG would like to thank E. Pérez for several informative conversations. EMV would like to thank M. Rodriguez-Meza for discussions and suggestions.

**Supplementary Information Available:** Full details of the DPD methodology, temperature parametrization, all density profiles at various temperature and additional data are provided. See DOI:10.1039/c4sm01981d